\documentclass[12pt,reqno]{amsart}
\allowdisplaybreaks[4]
\usepackage{graphicx} 
\usepackage{amssymb}
\usepackage{xcolor}
\usepackage{amsmath}
\usepackage{cite}
\usepackage{subfigure}
\usepackage{graphicx}
\usepackage{epstopdf}
\usepackage{bibentry}
\DeclareMathOperator{\sech}{sech}

\begin{document}

\author{L. J. Guo}

\author{C. B. Ward}
\thanks{Corresponding Author: ward@math.umass.edu}

\author{I. K. Mylonas}

\author{P. G. Kevrekidis}
\address{Department of Mathematics and Statistics, University of
	Massachusetts, Amherst MA 01003-4515, USA}

\title{Solitary Waves of the Camassa-Holm Derivative Nonlinear Schr\"odinger equation}

\begin{abstract}
In this paper we examine a deformation of the derivative nonlinear Schr\"odinger (DNLS) equation, the so-called Camassa-Holm DNLS (CH-DNLS)
equation. We use two asymptotic multiscale expansion methods to reduce this model to both the modified Korteweg-de Vries (MKdV) equation and the Korteweg-de Vries(KdV) equation. Using their exact soliton solutions, we construct approximate solutions of the original CH-DNLS equation, namely dark and anti-dark solitary waves. Direct numerical simulations demonstrate the validity of these approximate solutions and illustrate their  dynamical evolution, including
their potential for nearly elastic head-on collisions in the case
examples considered.
\end{abstract}

\maketitle
\section{Introduction}
The Casmassa-Holm (CH) equation\cite{PRL71} was proposed by Camassa and Holm as a nonlinear model for the unidirectional
propagation of shallow water waves over a flat bottom\cite{ARMA,JFM}. It is completely integrable in the sense that it admits a Lax pair, a bi-Hamiltonian structure, and an infinite number of conservation laws\cite{JDF,inverse,JPA38}.
One of the key characteristics of the equation~\cite{Inv,Bull,Ann}
is the presence of peaked traveling wave solutions
(the so-called peakons) that are strongly reminiscent of features
present in water waves.

Recently, Arnaudon\cite{JNS26} proposed a series of deformed equations corresponding to some classical soliton equations by developing a theory of Lagrangian reduction. Among these, the focusing Camassa-Holm Nonlinear Schr\"odinger (CH-NLS) equation was studied both analytically and numerically in Ref.\cite{JPA49}. Although its integrability remains an open problem, its solitary wave solutions describe some typical soliton properties including near elastic collisions.
In addition, the use of multiscale expansion methods\cite{1982},
allows one to  reduce integrable (and even non-integrable) models to
some prototypical integrable ones, as a means of characterizing the
solutions of the former over suitable (long) spatial and termporal
scales. For example, the reduction of certain perturbed defocusing NLS equations to the KdV model derived not only  approximate dark (gray) soliton solutions but also the prediction of novel structures such as the anti-dark solitons, having the form of humps (instead of dips) on top of a continuous wave background~\cite{PRA42,PRA44,JPA29}.
Recently, for the defocusing CH-NLS model~\cite{PLA381} (for which
analytical solutions have not been reported to the best of our knowledge),
some of the present authors analyzed its approximate dark and anti-dark
coherent structures by resorting to asymptotic multiscale expansion techniques and relying on the reduction of the CH-NLS model to the KdV equation.
Going beyond 1D systems, such methods have also been used to investigate the approximate line, ring and lump soliton solutions of a variety of
two-dimensional (2D)
systems~\cite{PLA285,OPT41,JPA492016,PRE64,PRL116,PRL99,PRL118,arxiv}.

On the other hand,
in recent years, the derivative NLS (DNLS) equation has attracted considerable attention both from the theoretical point of view and with respect
to physical applications. In plasma physics it has long been known that the DNLS equation governs the evolution of small but finite amplitude Alfv\'en waves propagating quasi-parallel to the magnetic field \cite{PS40,JPP16}. This equation is also used to describe large-amplitude magnetohydrodynamic (MHD) waves in plasmas\cite{JPP67}. Under special conditions, bright solitons, dark solitons, anti-dark solitons, breather solutions, as well as rogue waves have been obtained in Refs.\cite{JPA44,CNSNS19}.

Inspired by the relevance of the CH-NLS deformation, the
interest in its dynamical behavior, and also by the physical
relevance of DNLS type-models,
in this paper we explore a deformed version of the DNLS equation, i.e.,  the CH-DNLS equation, which was constructed using the Lagrangian deformation and loop algebra splittings in~\cite{chdnls}. Here, we employ asymptotic multiscale expansion techniques to derive the modified KdV (MKdV) and KdV equations from the CH-DNLS equation. This allows us to construct approximate solutions of this
model in the form of dark and anti-dark solitons. This paper is organized as follows: in Section 2, we present the model and examine the modulational
stability analysis of its homogeneous equilibria. In Sections 3 and 4, we use two different types  of multiscale expansion methods in order
to derive MKdV and KdV equations, respectively. We then use their explicit solutions to construct two types of approximate soliton solutions of the original CH-DNLS equation, namely dark and anti-dark solitons. In Section 5, direct numerical simulations illustrate the validity of the approximate solutions and demonstrate their dynamical evolution as well as their nearly elastic collisions
for the perturbatively constructed small amplitude solutions
considered.  Finally, in Section 6, we
summarize our findings and present our conclusions, as well as
suggest a number of directions for future study.

\section{Model and Its stability}
The CH-DNLS equation is given by
\begin{equation}\label{1}
\hat{q_t}=i q_{xx}+2((a^2|q_x|^2-|q|^2)\hat{q})_x, \quad \hat{q}=q-a^2q_{xx}, \quad a\in\mathbb{R},
\end{equation}
where $\hat{q}$ and $q$ are complex functions with respect to variables $x$ and $t$.  When $a=0$, this equation reduces to the DNLS equation.
In terms of the complex filed $q$, the CH-DNLS equation can be read as
\begin{equation}\label{2}
q_t-iq_{xx}+2(|q|^2q)_x-a^2q_{xxt}-2a^2(q|q_x|^2-a^2|q_x|^2q_{xx}+|q|^2q_{xx})_x=0.
\end{equation}

It is obvious that the constant function $q=q_0$ solves Eq.~(\ref{2}). To study the linear stability of this constant solution, we perturb the CH-DNLS with the ansatz
\begin{equation}
\tilde{q}=q_0+\epsilon p
\end{equation}
where $\epsilon$ is a small parameter and $p(x,t)$ satisfies the following linearized (around the background $q_0$) equation
\begin{equation}
p_t-ip_{xx}-2a^2q_0^2p_{xxx}-a^2p_{xxt}+2q_0^2(2p+p^*)_x=0.
\end{equation}
Looking for solutions of the form $p(x,t)=r(t)e^{i\kappa x}+s(t)e^{-i\kappa x}$, we can
reduce the aforementioned linear PDE to the $2\times2$ system of ODE's $d \eta/dt=iM\eta$, where $\eta=[r,s^*]^{T}$ and
\begin{equation}
M=\frac{1}{1+a^2\kappa^2}\left(\begin{array}{cc}
-2\kappa^3a^2q_0^2-4\kappa q_0^2-\kappa^2 & -2\kappa q_0^2\\
 -2\kappa q_0^2 & -2\kappa ^3a^2q_0^2-4\kappa q_0^2+\kappa^2
\end{array}\right).
\end{equation}
The dynamics of $\eta(t)$ are therefore determined by the eigenvalues of this matrix, given by
\begin{equation}\label{6}
\omega_{1,2}=\frac{(-2\kappa^2a^2q_0^2-4q_0^2\pm\sqrt{4q_0^4+\kappa^2})\kappa}{1+a^2\kappa^2}.
\end{equation}
Given that the two roots are both real,  the constant solution $q_0$ is always
spectrally stable.

It's also worth noting that in the long-wavelength limit (i.e. as $\kappa \to 0$), the phase velocity $\omega/\kappa$  becomes
\begin{equation}
|c_1|=2q_0^2, \quad \text{or} \quad |c_2|=6q_0^2.
\end{equation}
These correspond to the Alfv\'en $(c_1)$ and magnetosonic $(c_2)$ MHD velocities\cite{PS40}. In the following, the MKdV equation can be derived from Eq.~(\ref{1}) by employing multiscale expansions around the velocity $c_1$, whereas the KdV equation can be derived by expanding around $c_2$.

\section{Reductive derivation of MKdV equation from CH-DNLS equation}

Let us introduce the well-known Madelung transformation
(or amplitude-phase decomposition)
 \begin{equation}\label{8}
 q=\rho \exp (i\phi),\quad \kappa(x,t)=\frac{\partial \phi}{\partial x}, \quad \omega(x,t)=-\frac{\partial \phi}{\partial t},
 \end{equation}
  where $\rho$ and $\phi$ are real functions with respect to variables $x$ and $t$. Substituting this  into Eq.~(\ref{2}), the simplest nontrivial solution is the constant background $q_0$. To better underline the hydrodynamic origin of the
soliton solutions presented below, we first derive a real MKdV equation. We thus seek
solutions of Eq.~(\ref{2}) in the form of the following
asymptotic expansion
\begin{equation}\label{9}
\rho=q_0+\epsilon \rho_1(\xi,\tau)+\cdots,\quad \kappa=\epsilon \kappa_1({\xi,\tau})+\cdots,
\end{equation}
where
\begin{equation}
\xi=\epsilon(x-c t), \quad \tau=\epsilon^3 t, \nonumber
\end{equation}
and $0<\epsilon\ll1$ is a formal small parameter.

Substituting Eqs.~(\ref{8}) and (\ref{9}) into Eq.(\ref{2}) and using the fact that
 $$\frac{\partial \kappa}{\partial t}+\frac{\partial \omega}{\partial x}=0,$$
we separate the resulting equation into real and imaginary parts to obtain, up to $O(\epsilon^4)$:
\begin{align}
0&= \epsilon \big[ 0 \big] +\epsilon^2 \big[ (6 q_0^2-c) \rho_{1\xi}+q_0 \kappa_{1\xi} \big] +\epsilon^3 \big[ (-3 a^2 c q_0+2 a^2 q_0^3) \kappa_1 \kappa_{1\xi} +(6q_0^2-c) \rho_{2\xi}+\kappa_{1\xi} \rho_1
\nonumber\\
&+2 \kappa_1 \rho_{1\xi}+q_0 \kappa_{2\xi}+12 q_0 \rho_1 \rho_{1\xi}\big] \label{10} \\
&+O(\epsilon^4)
\nonumber \\[.5cm]
0&=\epsilon \big[ (2 q_0^3-c q_0) \kappa_{1\xi} \big] +\epsilon^2 \big[  (6 q_0^2-c) \kappa_{1\xi} \rho_1+(6 q_0^2-c) \kappa_1 \rho_{1\xi} +(2 q_0^3- c q_0) \kappa_{2\xi}+2 q_0 \kappa_1 \kappa_{1\xi} \big]
\nonumber \\
&+\epsilon^3 \big[ -3 a^2 c q_0 \kappa_1^2 \kappa_{1\xi}+(a^2 c q_0 -2 a^2 q_0^3) \kappa_{1\xi \xi \xi}-c \kappa_{1\xi} \rho_2-c \kappa_1 \rho_{2\xi}+(2 q_0 \kappa_1 -c \rho_1) \kappa_{2\xi}
\nonumber \\
&+(2 q_0 \kappa_{1\xi}-c \rho_{1\xi}) \kappa_2+(2 q_0^3-c q_0) \kappa_{3\xi}+6 q_0 \kappa_{1\xi} \rho_1^2+2 \kappa_1 \kappa_{1\xi} \rho_1+6 q_0^2 \kappa_{1\xi} \rho_2+q_0 \kappa_{1\tau}
\label{11}\\
&+12 q_0 \kappa_1 \rho_1 \rho_{1\xi}+\kappa_1^2 \rho_{1\xi}+
6 q_0^2 \kappa_1 \rho_{2\xi}+6 q_0^2 \kappa_{2\xi} \rho_1+6 q_0^2 \kappa_2 \rho_{1\xi}-\rho_{1\xi \xi \xi}\big]
\nonumber \\
&+O(\epsilon^4). \nonumber
\end{align}
Notice that to $O(\epsilon)$ Eq.~(\ref{11}) is satisfied by
\begin{equation}\label{12}
c=2q_0^2,
\end{equation}
as expected. Now, using this identity, Eqs.~(\ref{10}) and (\ref{11}) to $O(\epsilon^2)$ are zero provided
\begin{equation}\label{100}
\kappa_{1}=-4q_0\rho_1.
\end{equation}
Using the previous two equations, we see that Eq.~\eqref{10} is satisfied at $O(\epsilon^3)$ provided
\begin{equation}\label{13}
\kappa_2=32a^2q_0^4\rho_1^2-4q_0\rho_2
\end{equation}
Substituting Eqs.~(\ref{12}), (\ref{100}), and \eqref{13} into Eq.(\ref{11}), we see that it vanishes to $O(\epsilon^3)$ provided that $\rho_1$ satisfies the real MKdV equation
\begin{equation}\label{14}
4q_0^2\rho_{1\tau}+\rho_{1\xi\xi\xi}+24q_0^2\rho_1^2\rho_{1\xi}=0.
\end{equation}
Under the scaling transformation
\begin{equation}
 \hat{\xi}=2q_0\xi, \quad  \hat{\tau}=2q_0\tau,
 \end{equation}
Eq.~\eqref{14} becomes a standard MKdV equation
\begin{equation}\label{16}
\rho_{1\hat{\tau}}+\rho_{1\hat{\xi}\hat{\xi}\hat{\xi}}+6\rho_1^2\rho_{1\hat{\xi}}=0.
\end{equation}
This equation is well-known to admit a soliton solution
(see, e.g.,~\cite{Drazin,NDW})
\begin{equation}
\rho_1=b\sech[b(\hat{\xi}-\hat{\xi_0})-b^3\hat{\tau}]
\end{equation}
where  arbitrary constants $b$ and $\hat {\xi_0}$ determine the amplitude (and width and velocity) of the soliton. Returning to the original $(x,t)$-coordinates, we find the following approximate solution of the CH-DNLS equation, Eq.(\ref{2}):
\begin{equation}\label{MKdV Soliton}
\begin{aligned}
q\approx &\bigg [ q_0+\epsilon b\sech(\chi) \bigg] \exp\bigg[-2i\arctan(\sinh(\chi))\bigg],
\end{aligned}
\end{equation}
where
\[\chi=2bq_0\epsilon (x-x_0)-(4bq_0^3\epsilon+2b^3q_0\epsilon^3)t.\]

It's important to note that this perturbative solution describes two types of solitons: for $b>0$, the soliton is anti-dark, corresponding to density humps on top of the constant background $q_0$, whereas for $b<0$  the soliton is dark, corresponding to density dips on top of
the constant background $q_0$. 

{It should also be noted that although $\kappa$ has been found to $O(\epsilon)$ the phase itself is only given here to $O(1)$. The reason for this  is that when returning to the $x,t$ coordinates the $\epsilon$'s in the coordinate transformation cancel with the $\epsilon$'s in the expansion for the phase. This also occurs in the reduction of the DNLS equation to the MKdV equation.}

\section{Reductive Derivation of KdV Equation from CH-DNLS Equation}
We now proceed to derive the KdV equation from Eq.~(\ref{2}) by using a multiscale expansion method around the velocity $c_2$. First, rewrite Eq.~(\ref{2}) as two real equations by using the ansatz
\begin{equation}\label{20}
	q=q_{y}+i q_{z}.
\end{equation}
We then seek solutions in the form of the following asymptotic expansions:
\begin{eqnarray}\label{21}
	q_{y}=q_{0}+\epsilon q_{y}^{(1)}+\epsilon^2 q_{y}^{(2)}+... \nonumber\\
	q_{z}=\epsilon^{3/2}(q_{z}^{(1)}+\epsilon q_{z}^{(2)}+...)
	\nonumber\\
	\tau=\epsilon^{3/2}t , \quad \xi=\sqrt{\epsilon}(x-ct)
\end{eqnarray}
Substituting Eqs.~(\ref{20})-(\ref{21}) into Eq.~(\ref{2}), we obtain the following results:

\begin{align}
0&= \epsilon^{3/2} \big[ (6 q_0^2-c)q^{(1)}_{y\xi}\big]
\nonumber \\
&+\epsilon^{5/2} \big[ q^{(1)}_{y\tau}+q^{(1)}_{z\xi\xi}+(ca^2-2a^2q_0^2)q^{(1)}_{y\xi\xi\xi}+
12q_0q^{(1)}_{y}q^{(1)}_{y\xi}+(6q_0^2-c)q^{(2)}_{y\xi}\big]
\label{22}\\
&+ O(\epsilon^{7/2})
\nonumber\\[.4cm]
0&=\epsilon^2 \big[(2q_0^2-c)q^{(1)}_{z\xi}-q^{(1)}_{y\xi\xi}\big]
\nonumber \\
&+O(\epsilon^3) \label{23}
\end{align}
Eq.~(\ref{22}) to $O(\epsilon^{3/2})$ yields
\begin{equation}\label{24}
c=6q_0^2.
\end{equation}
Similarly, Eq.~(\ref{23}) to $O(\epsilon^{2})$ is satisfied by
\begin{equation}\label{25}
	q_{z}^{(1)}=-\frac{1}{4q_{0}^{2}}\frac{\partial q_{y}^{(1)}}{\partial \xi}.
\end{equation}
Now, substituting Eqs.~(\ref{24}) and (\ref{25}) into Eq.~(\ref{22}) at $O(\epsilon^{5/2})$ yields the following KdV equation for $q^{(1)}_y$
\begin{equation}\label{26}
	\frac{\partial q_{y}^{(1)}}{\partial \tau}+12q_{0}q_{y}^{(1)}\frac{\partial q_{y}^{(1)}}{\partial \xi}+(4a^2q_{0}^2-\frac{1}{4q_{0}^{2}})\frac{\partial^3 q_{y}^{(1)}}{\partial \xi^3}=0.
\end{equation}
To proceed further, and express the KdV equation (\ref{26}) in its standard form, we introduce
the straightforward rescaling:
\begin{eqnarray}
	\tilde{\tau}=\frac{16a^2q_0^4-1}{4q_0^2}\tau,\quad q_{y}^{(1)}=-\frac{16a^2q_0^4-1}{8q_0^3}\tilde{q}_{y}^{(1)}.
\end{eqnarray}
Then Eq.~(\ref{26}) is reduced to
\begin{equation}
\frac{\partial \tilde{q}_{y}^{(1)}}{\partial \tilde{\tau}}-6\tilde{q}_{y}^{(1)}\frac{\partial \tilde{q}_{y}^{(1)}}{\partial \xi}+\frac{\partial^3 \tilde{q}_{y}^{(1)}}{\partial \xi^3}=0.
\end{equation}
The above equation possesses the commonly known (see, e.g., Ref. \cite{Drazin,NDW}) soliton solution
\begin{equation}\label{27}
\tilde{q}_{y}^{(1)}=-\frac{\lambda}{2}\sech^{2}[\frac{\sqrt{\lambda}}{2}(\xi-\lambda\tilde{\tau}+\xi_{0})],
\end{equation}
with arbitrary constants $\lambda>0$ and $\xi_0$. Using this exact solution and returning to the original variables, we have that the perturbative approximation of Eq.~\eqref{2} is given by
\begin{equation} \label{mkdv28}
\begin{aligned}
&q= q_{y}+iq_{z},\\
&q_{y} \approx q_0-\epsilon \frac{\lambda a_2}{2}\sech^2(\chi),\\
&q_{z} \approx -\epsilon^{3/2} \frac{\lambda^{3/2} a_2}{8q_0^2}\text{sech}^2(\chi)\tanh(\chi),
\end{aligned}
\end{equation}
where parameters $\chi$, $a_1$, and $a_2$ are defined as
\begin{eqnarray}
\chi = \frac{\sqrt{\lambda\epsilon}}{2}[(x-x_0)-(6q_0^2+a_1\lambda\epsilon)t] , \\[.4cm]
a_1=\frac{16a^2q_0^4-1}{4q_0^2}, \quad a_2=-\frac{16a^2q_0^4-1}{8q_0^3}.
\end{eqnarray}
We see that this soliton can be either dark if $a_2 > 0$ or antidark if $a_2 < 0$. It is relevant to discuss here an interesting distinction between
the MKdV and KdV reduction results. In the former, the
potential dark or antidark nature of the solitonic structures
depends on the choice of the parameter $b$ controlling the (arbitrary)
amplitude, width, and velocity of the wave. On the other hand,
this type of freedom does {\it not} exist in the KdV reduction
whereby the nature of the wave is controlled by the height of
the background $q_0$ and the CH-deformation model parameter $a$.

\section{Numerically Obtained Solutions}

\begin{figure}[!htbp]
\centering
\subfigure[Antidark MKdV Soliton]{\includegraphics[width=.4\textwidth]{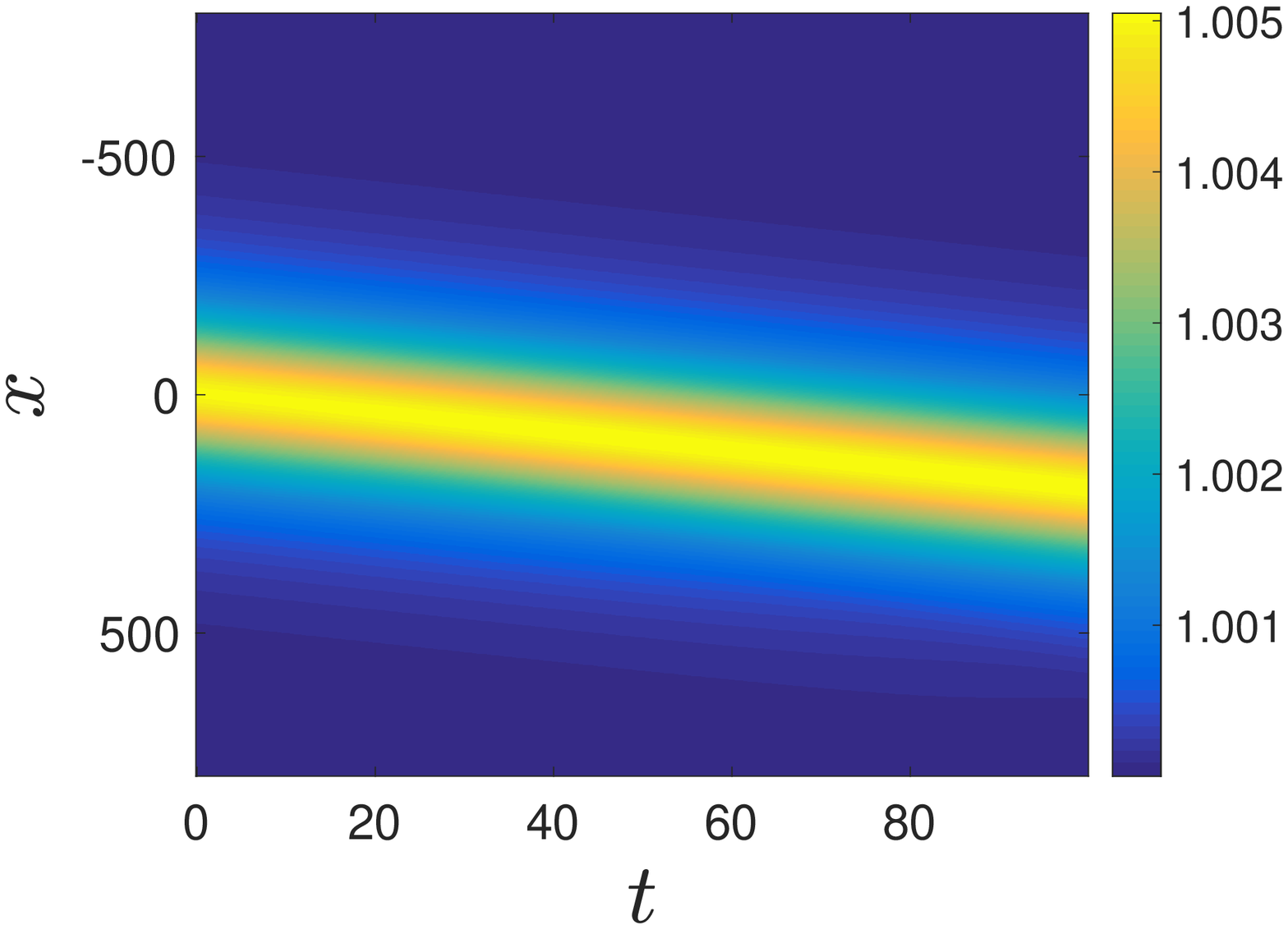}} \quad
\subfigure[Dark MKdV Soliton]{\includegraphics[width=.4\textwidth]{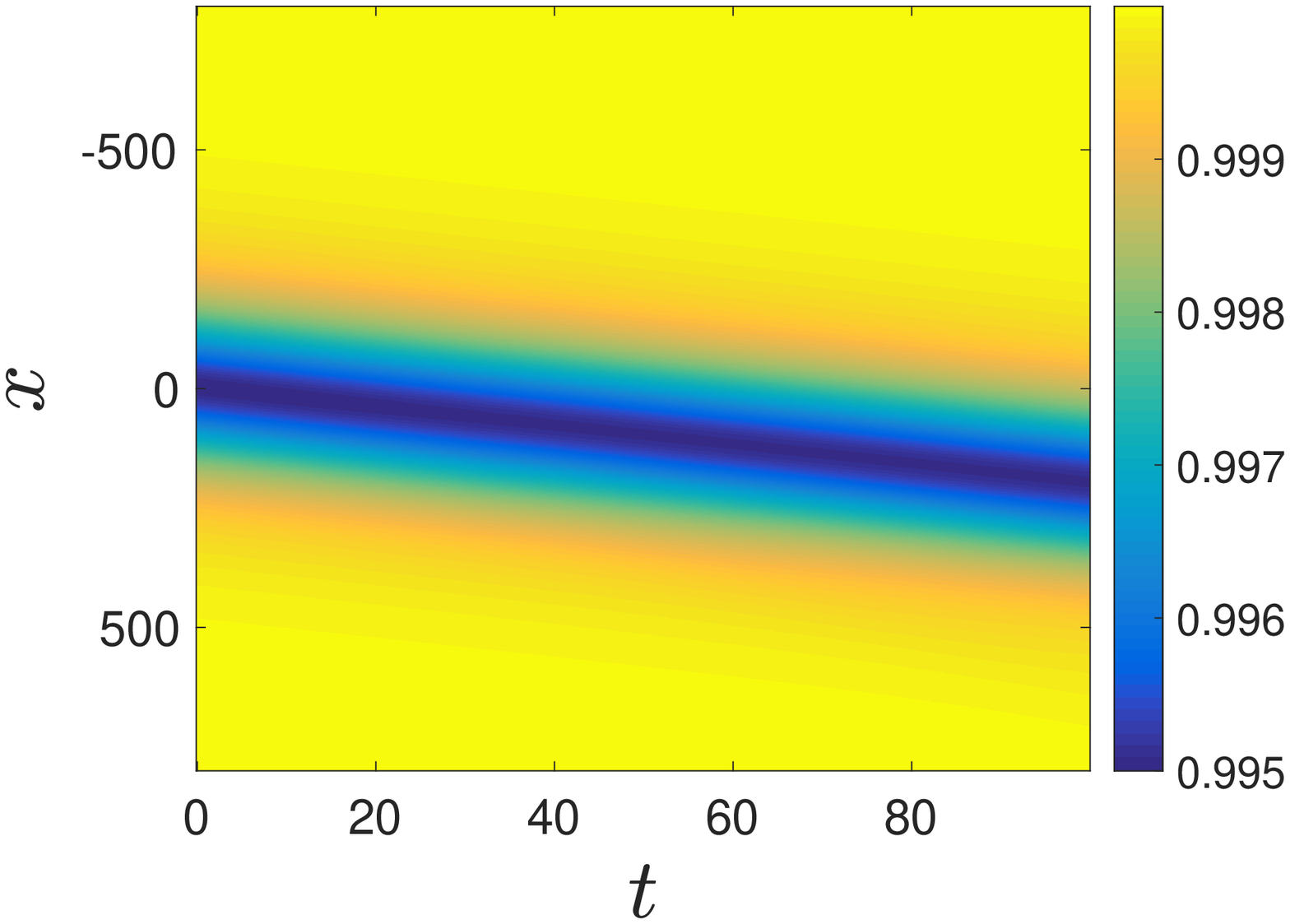}}

\subfigure[Antidark KdV Soliton]{\includegraphics[width=.4\textwidth]{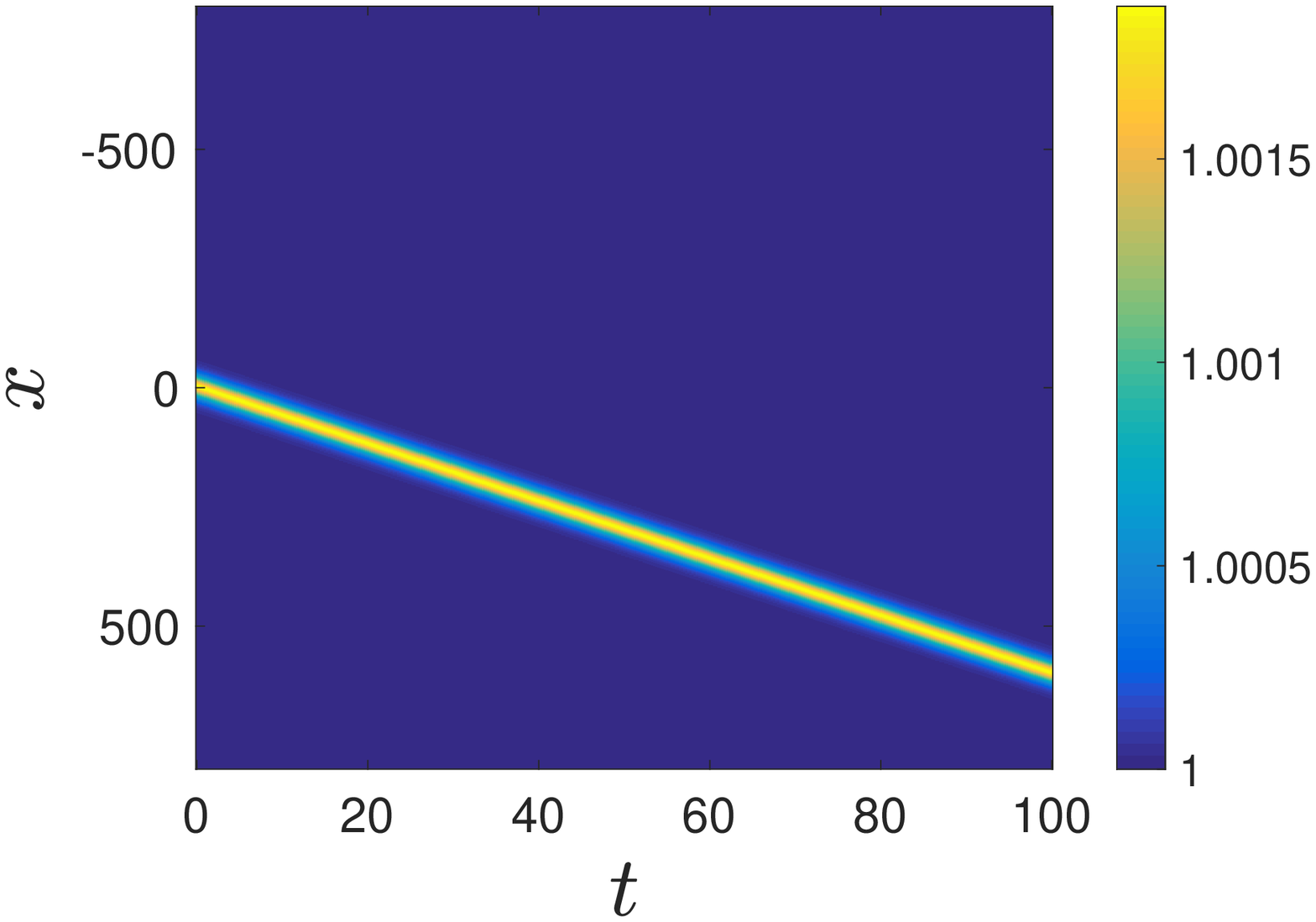}} \quad
\subfigure[Dark KdV Soliton]{\includegraphics[width=.4\textwidth]{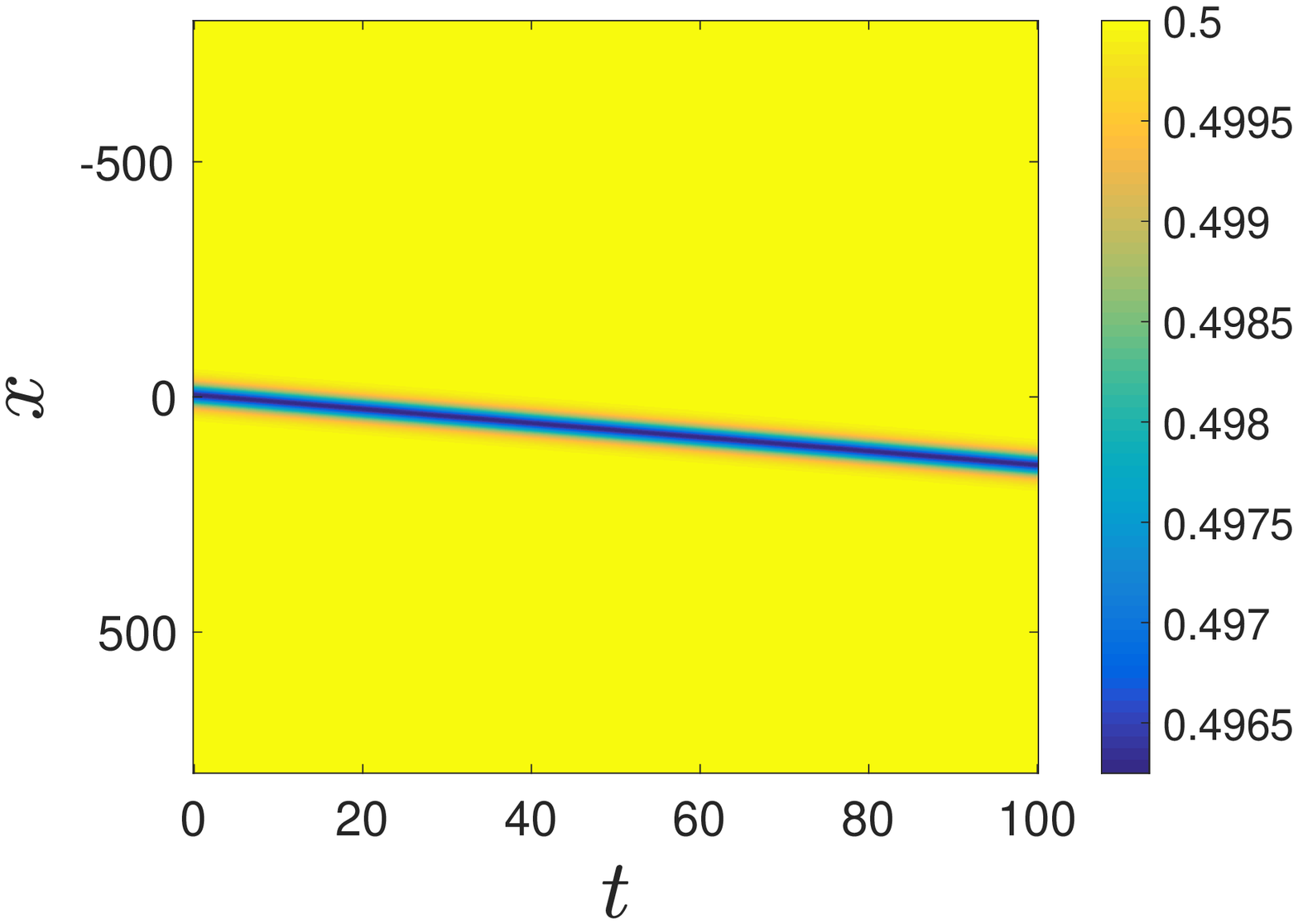}}
\caption{Small Amplitude Solitons. {Note that the speed of the KdV and MKdV solitons are nearly identitical to what the asympototic expansion predicts.} Parameter values used: (a) $\epsilon=0.005$, $a=0.5$, $b=-1$, $q_0=1$ (b) $\epsilon=0.005$, $a=0.5$, $b=1$, $q_0=1$ (c) $\epsilon=0.01$, $a=0.5$, $\lambda=1$, $q_0=1$ (d) $\epsilon=0.01$, $a=0.5$, $\lambda=1$, $q_0=0.5$.}\label{Fig1}
\end{figure}

\begin{figure}
\centering
\subfigure[Antidark MKdV Soliton]{\includegraphics[width=.4\textwidth]{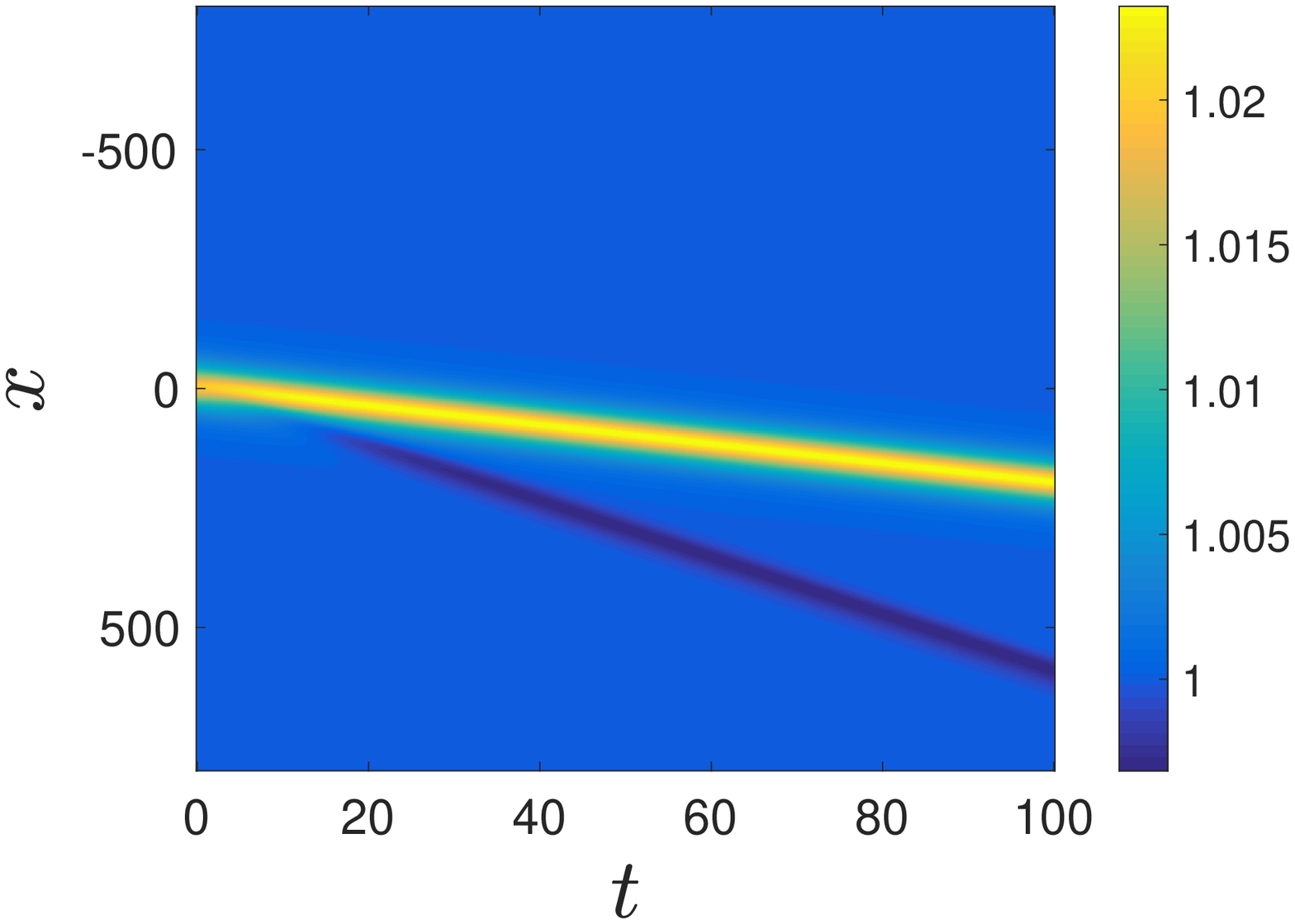}}
\quad
\subfigure[Dark MKdV Soliton]{\includegraphics[width=.4\textwidth]{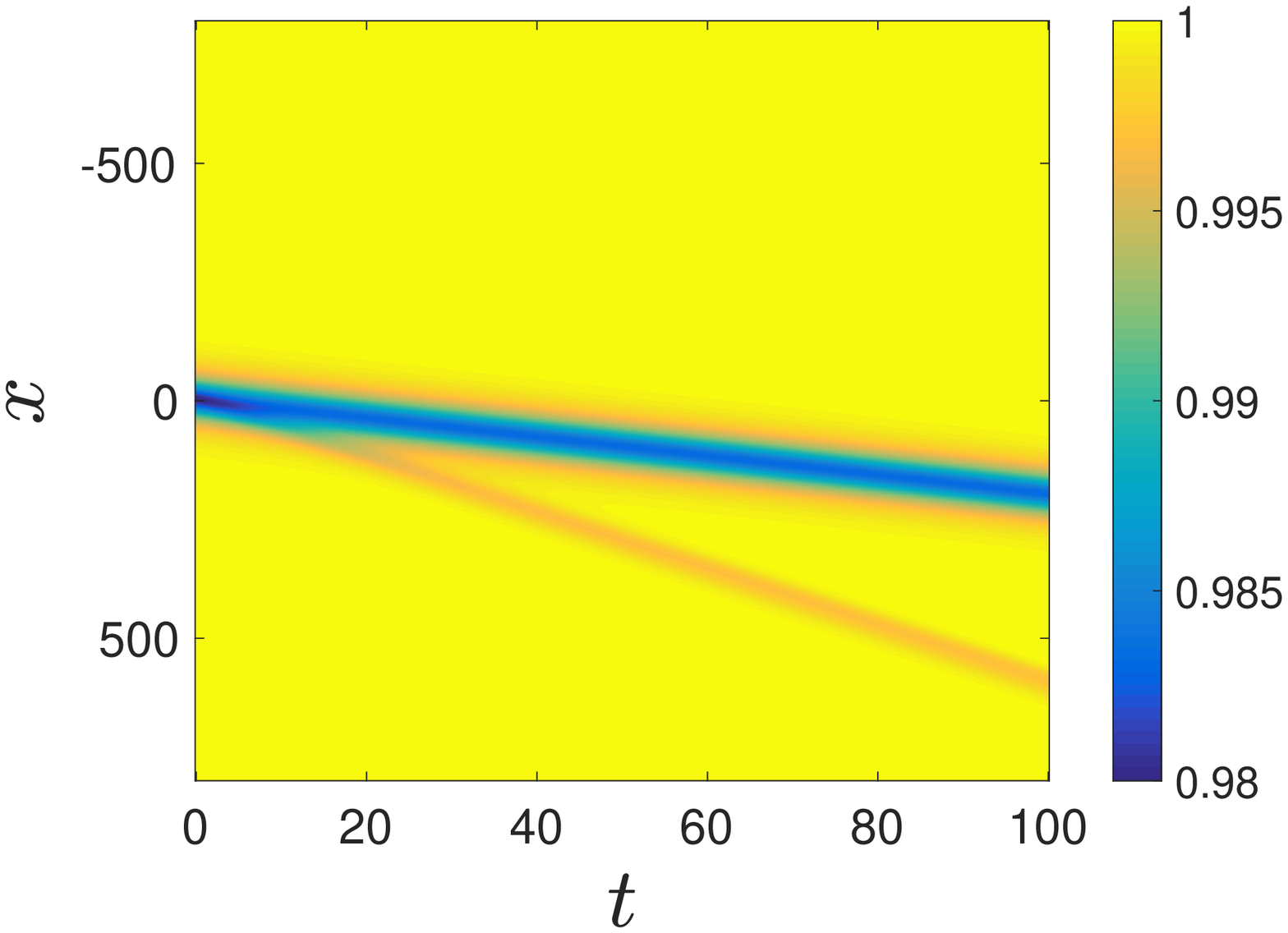}}

\subfigure[Antidark KdV Soliton]{\includegraphics[width=.4\textwidth]{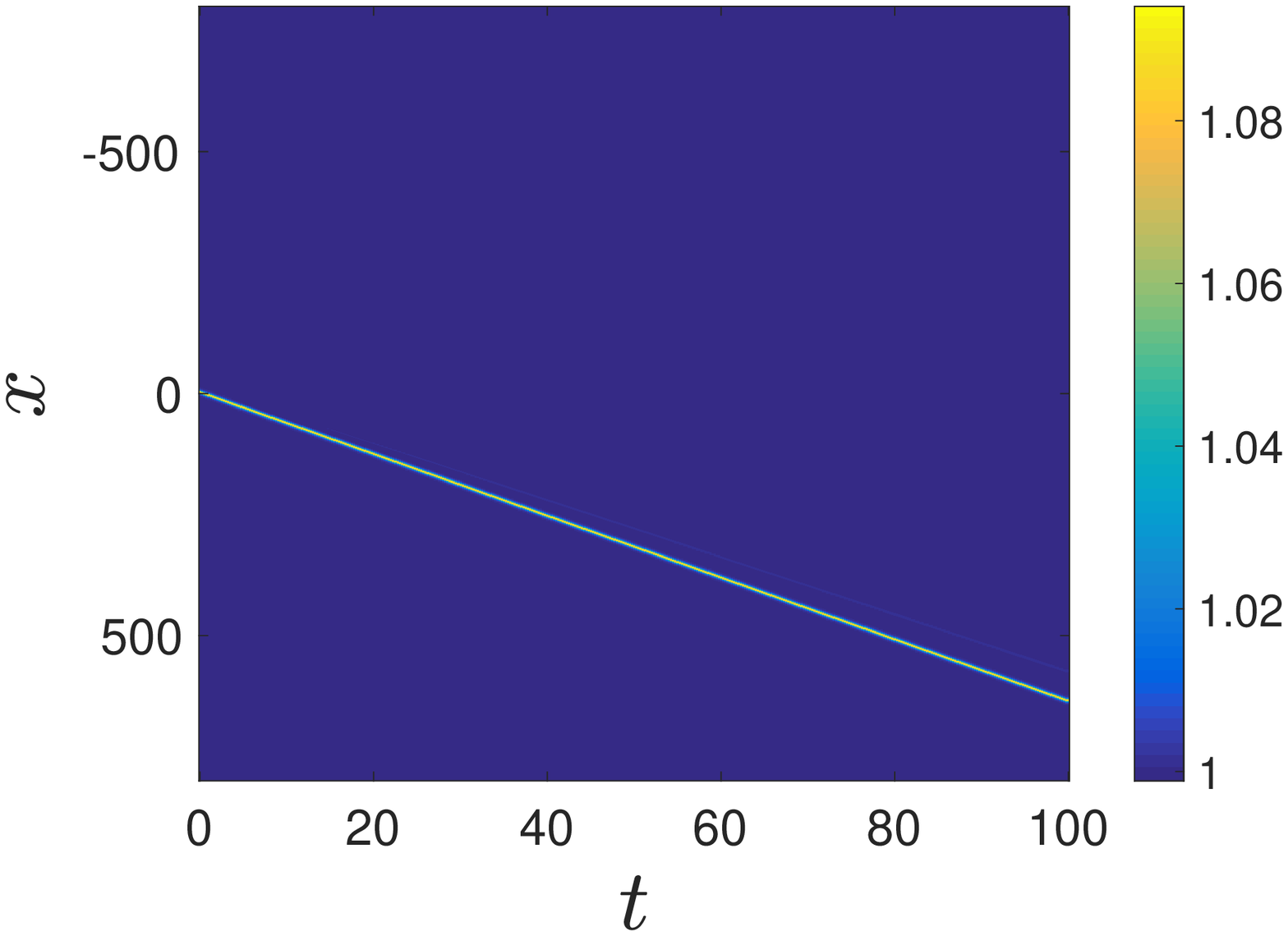}}
\quad
\subfigure[Dark KdV Soliton]{\includegraphics[width=.4\textwidth]{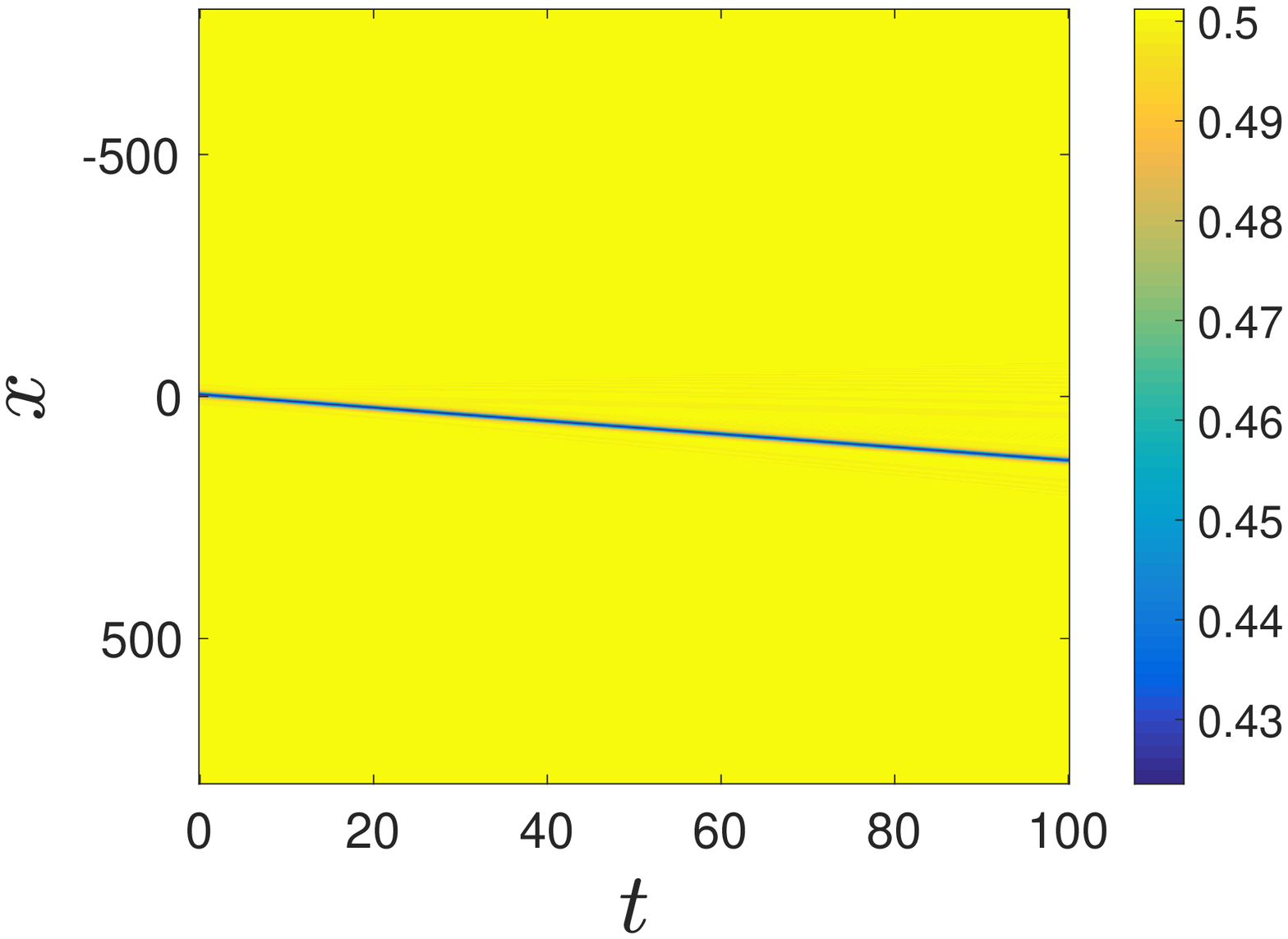}}
\caption{Large Amplitude Solitons. We see a similar qualitative behavior as in the small amplitude case, except now with some small radiation (as the
  solution initially ``adapts''). Parameter values used: (a) $\epsilon=0.02$, $a=1$, $b=-1$, $q_0=1$ (b) $\epsilon=0.02$, $a=1$, $b=1$, $q_0=1$ (c) $\epsilon=0.5$, $a=0.5$, $\lambda=1$, $q_0=1$ (d) $\epsilon=0.2$, $a=0.5$, $\lambda=1$, $q_0=0.5$.}\label{Fig2}
\end{figure}

In this section we corroborate our
analytical predictions by numerically integrating Eq.~(\ref{1})
using suitable initial and boundary conditions.
Our aim is to confirm the existence of the previously identified
asymptotic solutions and to explore these solutions at large amplitude
(i.e., in the regime where the asymptotic reduction leading to their
identification should not be expected to be valid).

\begin{figure}
\centering
\subfigure[Antidark KdV/Antidark MKdV Soliton]{\includegraphics[width=.4\textwidth]{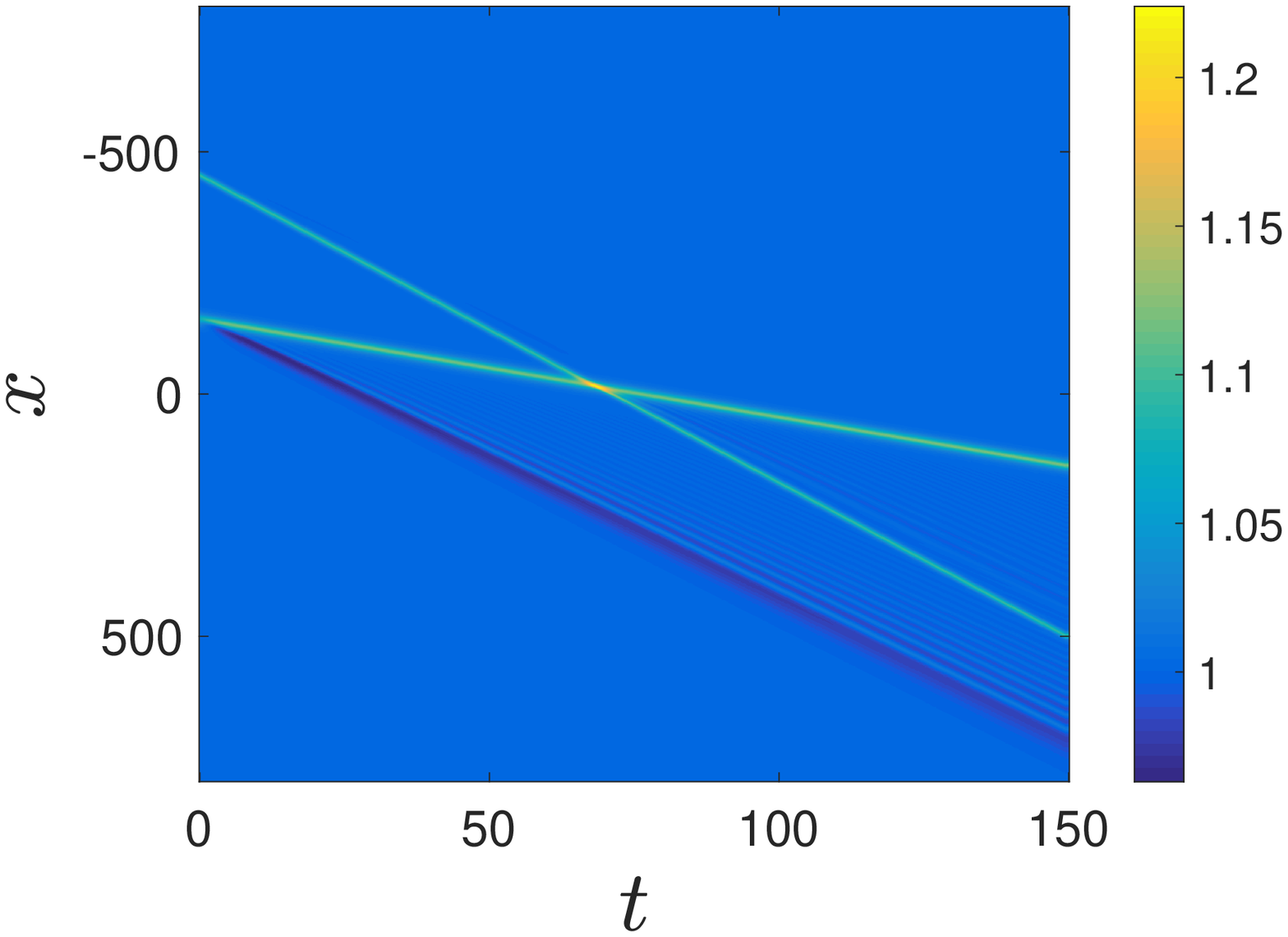}}
\subfigure[Antidark KdV/Dark MKdV Soliton]{\includegraphics[width=.4\textwidth]{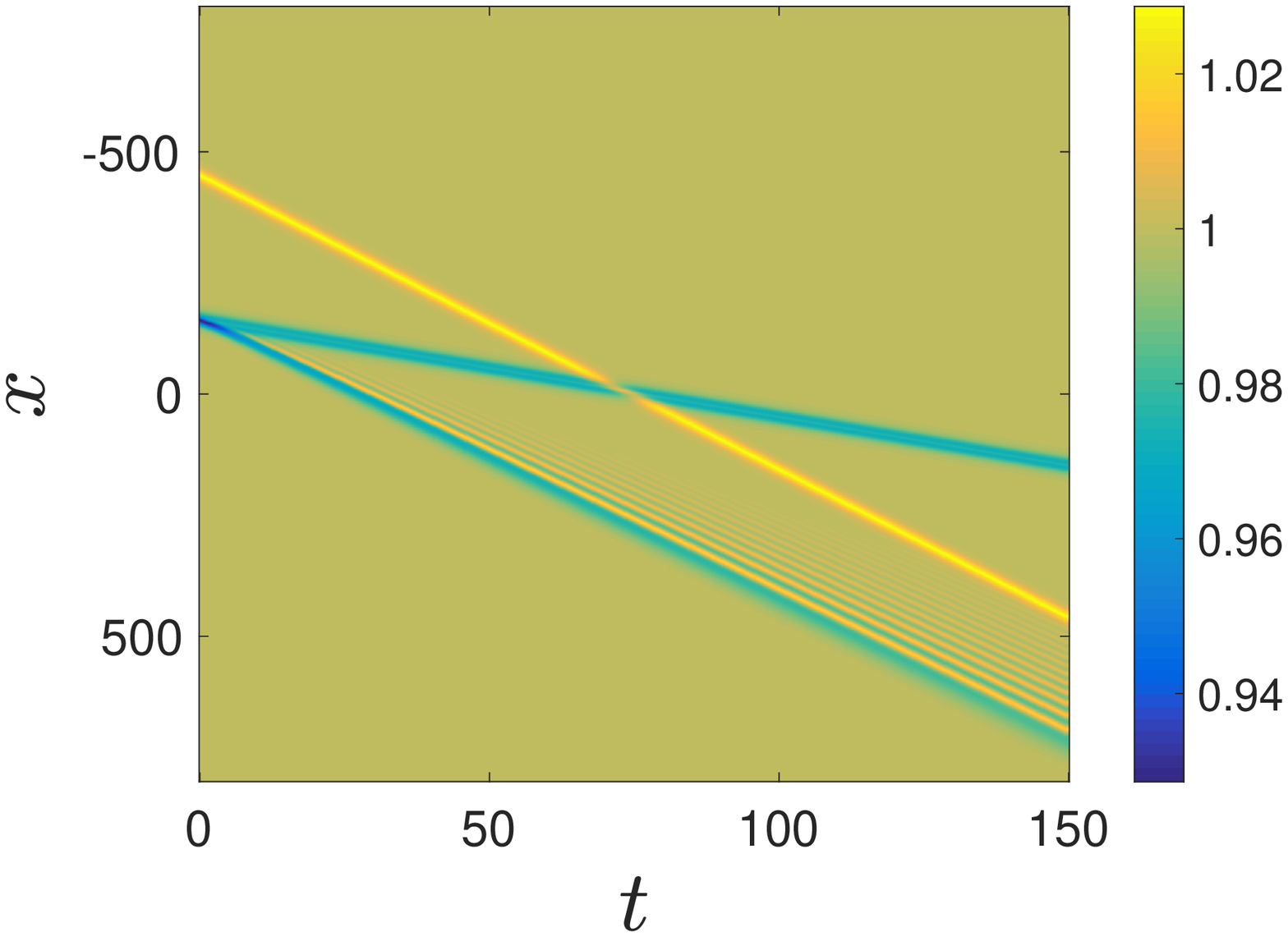}}
\caption{Large Amplitude Collisions. (a) $\epsilon_{\text{KdV}}=0.1$, $\lambda=1$, $\epsilon_{\text{MKdV}}=0.07$, $b=-1$, $a=1$, $q_0=1$ (b) $\epsilon_{\text{KdV}}=0.03$, $\lambda=1$, $\epsilon_{\text{MKdV}}=0.07$, $b=1$, $a=1$, $q_0=1$. We can see that
  the antidark KdV-antidark MKdV, as well as the antidark KdV-dark MKdV
  collisions are nearly elastic for the considered parameters and initial
conditions.}\label{Fig3}
\end{figure}

Fig.~\ref{Fig1} corresponds to the case of small amplitudes while Figs.~\ref{Fig2} and \ref{Fig3} correspond to large amplitudes. In order to be
more precise, let us note that 
for the initial condition of the KdV soliton, we set $t$ equal to zero in Eqs.~\eqref{mkdv28}. We then used the ansatz
\begin{equation}\label{Eq12}
q_i(x) :=\bigg(q(x,0)-q_0\bigg)\exp\bigg[-\bigg(\frac{x}{L^*}\bigg)^{12}\bigg]+q_0
\end{equation}
as the initial condition, where $q(x,0)$ is as given in Eqs.~\eqref{mkdv28}. Here $L^*$ is set to 0.8 times the length of the (spatial) computational domain. This ensures that the the initial condition $q_i$ is, to numerical precision,
equal to $q_0$ at the boundaries, allowing for the use of periodic boundary conditions. Further, $L^*$ is chosen large enough so that the boundaries do not effect the interior dynamics during the time interval of the simulation.
In a similar vein, for the MKdV solutions, we set $t$ equal to zero in Eqs.~\eqref{MKdV Soliton}, and then use the ansatz
\begin{equation}\label{Eq13}
q_i(x) :=\bigg(q(x,0)+q_0\bigg)\exp\bigg[-\bigg(\frac{x}{L^*}\bigg)^{12}\bigg]-q_0
\end{equation}
as the initial condition. Due to the nature of the phase factor in Eq.~\eqref{MKdV Soliton}, the MKdV solitons asymptote to $-q_0$ for large values of $x,t$; hence, the ansatz Eq.~\eqref{Eq13} makes sure the value at the boundaries is indeed $-q_0$.

For collisions between the solitons, as an initial condition we simply multiplied the two previously defined initial conditions (i.e. the two ansatz corresponding to Eqs.~\eqref{Eq12} and \eqref{Eq13}); we've set the background $q_0=1$
for the KdV soliton and $-1$ one for the MKdV soliton, ensuring that they have the same value at far-field. We should mention, however, that we treat the $\epsilon$'s appearing in each ansatz as independent parameters; that is to say, we set $\epsilon$ appearing in the KdV soliton to some value $\epsilon_{KdV}$ and set $\epsilon$ appearing in the MKdV solitons to some other value $\epsilon_{MKdV}$.
{Finally, because the dark and antidark MKdV solitons have approximately identical speed, we did not consider collisions between the two,
  as they might require an extremely long interval of time integration
  for such an overtaking collision. Collisions between dark and antidark KdV solitons are not included either because the two types of solitons are not predicted to exist for the same value of the deformation parameter $a$.}

For the small amplitude time-integrations appearing in Fig.~\ref{Fig1} we see that the numerics and the theory are in very good agreement. In particular, notice that the predicted speed of the KdV solitons $6q_0^2=6$, $6q_0^2=1.5$ in panels (c),(d) and that of the MKdV solitons $2q_0^2=2$ in panels (a), (b) are seen to be almost identical to that given by the simulation. Moreover, the
coherent structures propagate essentially undistorted with these speeds,
as predicted by the reductive perturbation theory.

Fig.~\ref{Fig2} shows the results of the large amplitude initial conditions. As before, these are in good qualitative agreement with what the perturbation analysis suggests. It's worth pointing out that the ejected soliton appearing in Fig.~\ref{Fig2} (a),(b) is seen to not only have a speed of approximately 6 but it is also a \emph{dark} soliton, in contrast to our antidark KdV solitons which travel at the same speed.  

Fig.~\ref{Fig3} shows the results of colliding the antidark KdV and MKdV solitons. As can be seen, the solitons appear to collide nearly elastically, though radiation can clearly be seen to be created at the collision point in Fig.~\ref{Fig3}. This is especially visible in the left panel illustrating the collision of an antidark soliton of the KdV with one of the MKdV. Although in this case
too, the most substantial radiation arises from the ``adjustment'' of the
initial condition.

\section{Conclusion and Future Work}

In this work we have studied a deformation of the DNLS equation, the so-called CH-DNLS equation.
Given the wide applicability of DNLS equations in plasma physics and
magnetohydrodynamics and the relevance of CH-type deformations in contexts
associated with water waves, this is an interesting model to study.
The relevant partial differential equation
was reduced to the MKdV and KdV equations by employing two different multiscale expansion methods. The reduction to the MKdV and KdV equations allowed us to construct approximate solitary wave solutions of the original CH-DNLS equation by using the explicitly known  soliton solutions of the former.

We then used direct numerical simulation to demonstrate the relevance
of these solitary waves in the original CH-DNLS equation. Their dynamical evolution and their interactions were also analyzed and discussed. We found that for the small amplitude initial conditions(see Fig.~\ref{Fig1}), the numerics and the theory are in very good agreement. For larger amplitudes, the derived solutions persist and can undergo nearly elastic
head-on collisions, although some radiation can clearly be discerned
both at their initialization as well as during their collisions.

Numerous open questions still remain in this context. Whether one
can find exact solutions of the CH-DNLS model is of interest in its own
right. The dynamics of the DNLS equation (in higher
dimensions or {different powers of the nonlinearity}) is intriguing in connection
to collapse type features that have been of considerable recent
interest~\cite{simpson}. Examining how these features are modified
in the CH-DNLS case would be relevant to consider. 

{\bf Acknowledgments} { This work is supported by the NSF of China under Grant No. 11671219,  and the K.C. Wong Magna Fund in Ningbo University. L.J.G. also gratefully acknowledges China Scholarship Council for support of her study at the University of Massachusetts, Amherst. This material is based upon work
  supported 
by the National Science Foundation under Grant No. PHY-1602994 (P.G.K).}

\end{document}